WILEY-VCH

# Spin-charge interconversion in KTaO₃ two-dimensional electron gases

*Luis M. Vicente-Arche, Julien Bréhin, Sara Varotto, Maxen Cosset-Cheneau, Srijani Mallik, Raphaël Salazar, Paul Noël, Diogo Castro Vaz, Felix Trier, Suvam Bhattacharya, Anke Sander, Patrick Le Fèvre, François Bertran, Guilhem Saiz, Gerbold Ménard, Nicolas Bergeal, Agnès Barthélémy, Hai Li, Chia-Ching Lin, Dmitri E. Nikonov, Ian A. Young, Julien Rault, Laurent Vila, Jean-Philippe Attané and Manuel Bibes\**

Luis. M. Vicente-Arche, Julien Bréhin, Dr. Sara Varotto, Dr. Srijani Mallik, Dr. Diogo Vaz♣, Dr. Felix Trier♥, Suvam Bhattacharya♦, Dr. Anke Sander, Prof. Agnès Barthélémy & Dr. Manuel Bibes
Unité Mixte de Physique, CNRS, Thales, Université Paris-Saclay, 1 avenue Augustin Fresnel, 91767, Palaiseau France
E-mail : manuel.bibes@cnrs-thales.fr
♣ Present address: CIC nanoGUNE BRTA, Tolosa Hiribidea, 76, 20018 Donostia - San Sebastian, Spain
♥ Present address: Department of Energy Conversion and Storage, Technical University of Denmark, Fysikvej, Building 310, 2800 Kgs. Lyngby, Denmark
♦ Present address: Laboratoire de Physique des Solides, CNRS, Université Paris-Saclay, 91405 Orsay, France

Maxen Cosset-Cheneau, Dr. Paul Noël§, Dr. Laurent Vila, Dr. Jean-Philippe Attané
Université Grenoble Alpes, CEA, CNRS, Grenoble INP, SPINTEC, 38000 Grenoble, France
§ Present address : Dept. of Materials, ETH Zürich, Hönggerbergring 64, 8093 Zürich, Switzerland

Raphaël Salazar, Dr. Patrick Le Fèvre, Dr. François Bertran, Dr. Julien Rault
Synchrotron SOLEIL, L'Orme des Merisiers, Saint-Aubin, BP 48, 91192 Gif-sur-Yvette Cedex, France

Guilhem Saiz, Gerbold Ménard, Nicolas Bergeal
Laboratoire de Physique et d'Etude des Matériaux, ESPCI Paris, Université PSL, CNRS, Sorbonne Université, Paris, France.

Hai Li, Chia-Ching Lin, Dmitri Nikonov, Ian Young
Components Research, Intel Corp., Hillsboro, OR 97124, USA

Keywords: oxide interfaces, spin-orbit coupling, Edelstein effect, spin-charge interconversion

## Abstract

Oxide interfaces exhibit a broad range of physical effects stemming from broken inversion symmetry. In particular, they can display non-reciprocal phenomena when time reversal symmetry is also broken, e.g., by the application of a magnetic field. Examples include the direct and inverse Edelstein effects (DEE, IEE) that allow the interconversion between spin currents and charge currents. The DEE and IEE have been investigated in interfaces based on the perovskite $SrTiO_3$ (STO), albeit in separate studies focusing on one or the other. The demonstration of these effects remains mostly elusive in other oxide interface systems despite their blossoming in the last decade. Here, we report the observation of both the DEE and IEE in a new interfacial two-dimensional electron gas (2DEG) based on the perovskite oxide $KTaO_3$. We generate 2DEGs by the simple deposition of Al metal onto $KTaO_3$ single crystals, characterize them by angle-resolved photoemission spectroscopy and magnetotransport, and demonstrate the DEE through unidirectional magnetoresistance and the IEE by spin-pumping experiments. We compare the spin-charge interconversion efficiency with that of STO-based interfaces, relate it to the 2DEG electronic structure, and give perspectives for the implementation of $KTaO_3$ 2DEGs into spin-orbitronic devices.

The interconversion between spin and charge currents is an active research direction in spintronics exploiting the spin-orbit interaction in a variety of materials and heterostructures[1]. Spin-charge interconversion can be achieved by the direct and inverse spin Hall effects in bulk materials such as Pt or Ta or by their two-dimensional equivalent[2,3], but also through the direct and inverse Edelstein effects (DEE and IEE) in systems with substantial spin-orbit coupling and broken inversion symmetry. Interfaces between heavy metals (harboring a Rashba state), 2D materials or surfaces of topological insulators satisfy this condition and have been harnessed for spin-charge interconversion through DEE and IEE[4–7].

In 2016, Lesne *et al.* demonstrated a very large IEE in the two-dimensional electron gas (2DEG)[8] appearing at the interface between LaAlO$_3$ (LAO) and SrTiO$_3$ (STO)[9] for which a sizeable Rashba spin-orbit coupling had been identified[10]. By using spin-pumping ferromagnetic resonance (SP-FMR), they injected a spin current from a ferromagnetic layer of NiFe into the 2DEG and collected the transverse current generated by the IEE (cf. **Figure 1a-b**). Remarkably, the conversion figure of merit $\lambda_{IEE}$ (the inverse Edelstein length[4]) showed a strong dependence with gate voltage (and thus carrier density), even with a change of sign[8]. Subsequent experiments on 2DEGs, formed by depositing a reducing agent such as Al on STO, revealed that this variation of $\lambda_{IEE}$ with carrier density is linked to their rich multi-orbital band structure, with trivial and topological avoided crossings[11,12]. Fewer experiments have focused on charge-spin conversion in STO 2DEGs. Wang *et al.* used spin-torque ferromagnetic resonance (ST-FMR) and found a large efficiency at room temperature[13]. More recently, Choe *et al.* and Vaz *et al.* measured a unidirectional magnetoresistance (coined the bilinear magnetoresistance) in an LAO//STO 2DEG[14,15], caused by the DEE, which generates a spin density transverse to the applied charge current that changes sign when the current is reversed (see Figure 1c-d).

Another promising oxide system with potential for spin-charge interconversion is KTaO$_3$ (KTO). Similarly to STO, KTO is an incipient quantum ferroelectric[16] that becomes metallic when doped n-type with oxygen vacancies[17] for example, and can harbor a 2DEG at its surface or when interfaced with various materials[18,19]. As in STO, the deposition of a perovskite oxide film or the formation of oxygen vacancies (e.g., by ion irradiation[20]) can generate a 2DEG in KTO[21,22], possessing high mobilities and exhibiting signatures of spin-orbit coupling in low-temperature magnetoresistance data (i.e. weak antilocalization[19,23]). Because Ta is a 5d element and heavier than Ti, KTO is indeed expected to possess a larger spin-orbit coupling. Yet, to date, only one study has explored spin-charge conversion into KTO 2DEGs, which involved thermal spin injection from a ferromagnetic EuO overlayer with a modest current produced (~1 nA at 10 K)[24].

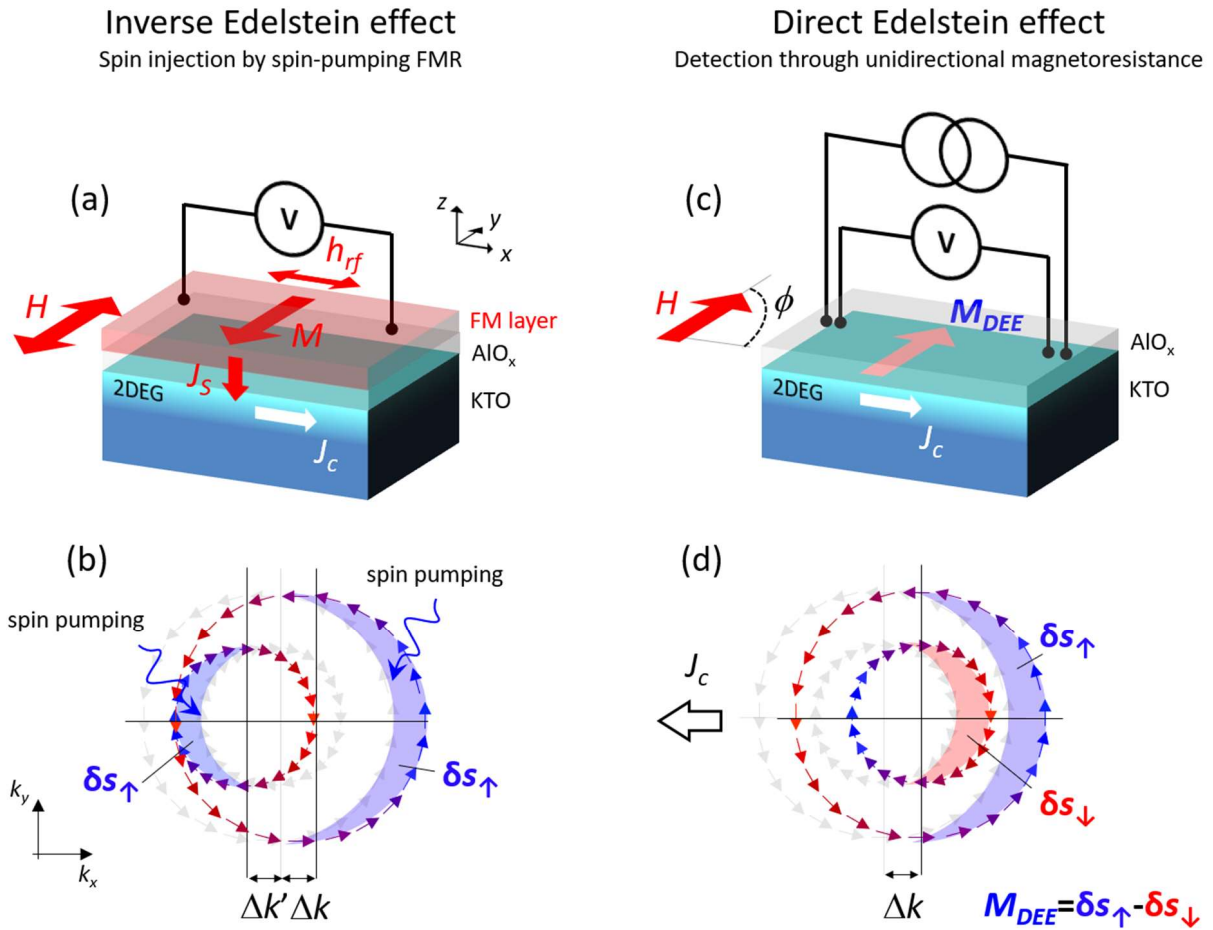

**Figure 1.** (a) Detection of the inverse Edelstein effect: when a dc magnetic field $H$ and a transverse radiofrequency magnetic field $h_{rf}$ induce ferromagnetic resonance in the ferromagnet with magnetization $M$, a spin current $J_S$ is injected by spin-pumping from the ferromagnet into the adjacent Rashba 2DEG. A transverse charge current $J_C$ is then generated through the IEE, producing under open circuit conditions a voltage $V$. (b) Physical mechanism: when the spin current is injected into the Rashba 2DEG, it produces a spin accumulation $\delta s_\uparrow$ in the two

Rashba-split Fermi contours with opposite spin chiralities, which causes them to shift along $k_x$ in opposite directions (by $\Delta k$ and $\Delta k' > \Delta k$), thereby generating a finite charge current along $x$. (c) Detection of the direct Edelstein effect: passing a charge current into the Rashba 2DEG produces a transverse spin density, i.e., a magnetization $M_{DEE}$, which generates a unidirectional magnetoresistance when the magnetic field is applied transverse to the current, that is parallel or antiparallel to $M_{DEE}$. Experimentally, we detect this effect by measuring the longitudinal resistance while rotating the magnetic field in the plane by an angle $\phi$ with respect to the current $J_C$. (d) Physical mechanism: the application of a current $J_C$ in the Rashba 2DEG causes a shift $\Delta k$ of the Fermi contours and produces spin accumulations $\delta s \uparrow$ and $\delta s \downarrow$ in each contour, which do not compensate and result in the generation of a finite spin density oriented along $y$.

Here we report the generation of 2DEGs in KTO by the deposition of ultrathin Al films at room temperature and their spin-charge interconversion properties. We study the 2DEG formation by performing X-ray photoelectron spectroscopy (XPS) *in situ* for increasing Al thickness, and confirm it by magnetotransport measurements. Angle-resolved photoemission (ARPES) data also reveal the presence of a 2DEG with a multiband structure, compatible with earlier results on KTO surfaces[25,26]. In samples covered with a NiFe layer we then carry out SP-FMR and observe spin-charge conversion with an efficiency comparable to that of LAO//STO 2DEGs. Finally, we probe charge-spin conversion by performing angle-dependent transport measurements and identifying a unidirectional magnetoresistance term, ascribed to the DEE-driven bilinear magnetoresistance effect. We extract an estimate of the Rashba coefficient and compare the results from both experiments and with existing data for STO interfaces.

**Figure 2a-c** shows the result of *in situ* XPS of the Ta $4f$ levels of KTO before (Fig. 2a) and after (Fig. 2b-c) the deposition of Al by sputtering. For the virgin KTO substrate, the spectrum can be well fitted by two components corresponding to $Ta^{5+}$ ions, as expected from stoichiometry. After the deposition of 10 Å of Al (Fig. 2b), some spectral weight appears at low binding energies, reflecting the reduction of $Ta^{5+}$ into $Ta^{4+}$ and $Ta^{2+}$. This indicates the population of the Ta $5d$ levels and points to the generation of an electron gas. The relative fraction of these reduced species increases for 21 Å of Al (Fig. 2c). Similar measurements were performed for additional Al thicknesses and the results are displayed in Fig. 2d. The relative

fractions of Ta$^{4+}$ and Ta$^{2+}$ increase with Al thickness at the expense of the amount of Ta$^{5+}$, indicating that the density of electrons in the Ta $5d$ bands increases for thicker Al. XPS of the Al $2p$ levels for 21 Å of Al showed that the Al was almost fully oxidized after exposing the samples to the atmosphere, consistent with results on Al//STO[12] (see Supplementary Material).

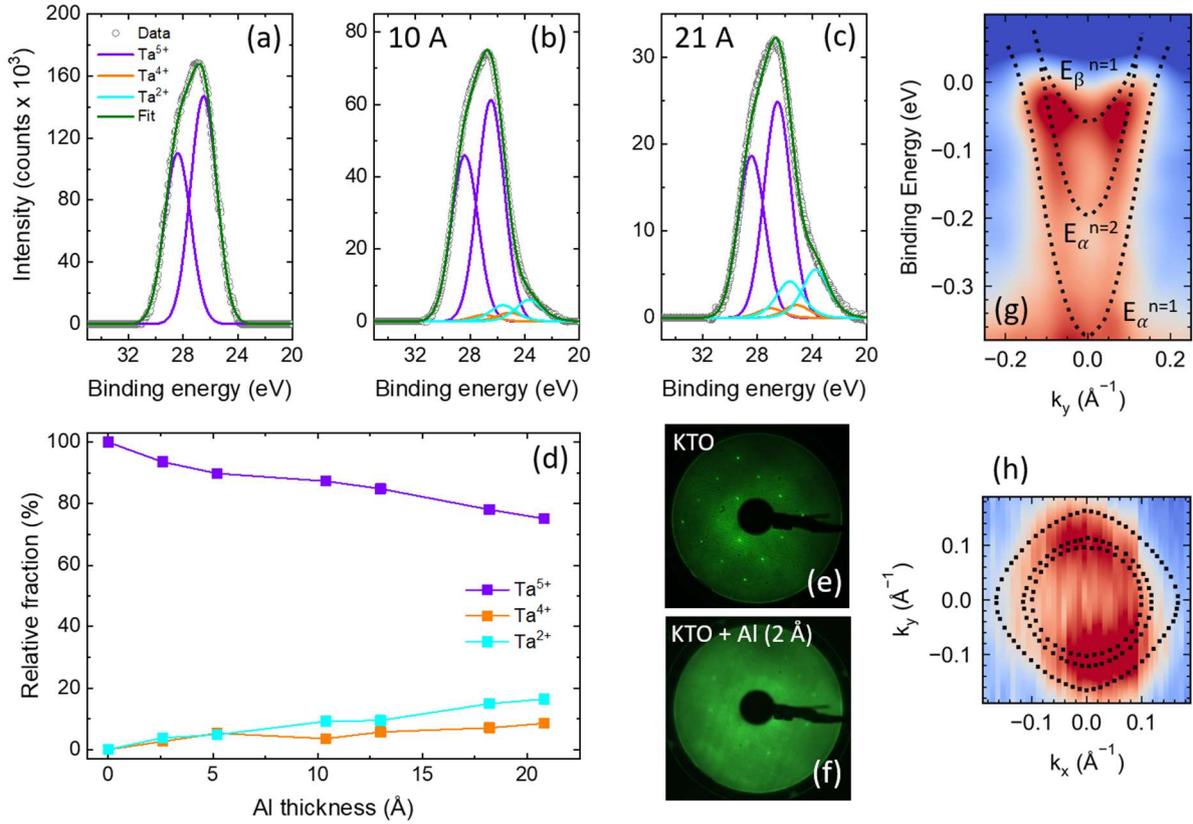

**Figure 2.** (a) XPS spectrum of the Ta $4f$ levels for a virgin KTO substrate. (b) XPS after deposition of 10 Å of Al. (c) XPS after deposition of 21 Å of Al. The data in (a-c) are shown as symbols and lines correspond to fit components and to their sum (green line). (d) Dependence of the relative fraction of the different Ta valence states with the Al thickness. LEED pattern for a bare KTO surface (e) and after deposition of 2 Å of Al (f). Dispersion curves (g) and corresponding Fermi surface (h) measured by ARPES for a Al(2 Å)//KTO sample. The dotted lines highlight the positions of the bands, following the results of Ref. [25]. The bands are labelled following the convention in Ref. [25].

Complementary photoemission experiments were performed at the Cassiopée beamline of Synchrotron SOLEIL. Al//KTO samples were prepared by growing a film of Al in a molecular beam epitaxy chamber that is connected under ultra-high vacuum to a LEED (low-energy electron diffraction) setup and to the ARPES chamber. Fig. 2e and 2f show LEED diffraction patterns of a KTO substrate surface before and after the deposition of 2 Å of Al.

Sharp diffraction spots corresponding to a square lattice attest to the high structural coherence of the surface. Fig. 2g presents band dispersions near $\Gamma_{002}$ probed by ARPES, which resemble those measured for KTO(001) surfaces[25,26] but without clear signs of the low dispersion band due to F centers mentioned in Ref.[25]. Measurements at different photon energies (not shown) confirmed that such electronic states are confined to the interface, as they do not disperse with $k_z$. Three parabolic bands are visible. The dotted lines indicate the positions of the bands detected in Ref. [25] which have been rescaled to better match our results. The bottom band (labelled $E_\alpha^{n=1}$, cf. [25]) has an effective mass of $m^* \approx 0.23 m_0$, the second band ($E_\alpha^{n=2}$) has $m^* \approx 0.23 m_0$, and the third band ($E_\beta^{n=1}$) has a mass $m^* \approx 0.52 m_0$, where $m_0$ is the free electron mass. Recent calculations suggest that $E_\beta^{n=1}$ should display a substantial Rashba splitting ($\alpha_R \approx 300$ meV.Å)[23], but this feature cannot be resolved here and was not detected in previous ARPES experiments[25,26]. The corresponding energy cuts at the Fermi level are shown in Fig. 2h, which show evidence of pseudo-circular Fermi surfaces (FSs). The dotted lines correspond to the FSs adapted from Ref. [25]. From the ARPES data, we estimate that the carrier densities of bands $E_\alpha^{n=1}$, $E_\alpha^{n=2}$ and $E_\beta^{n=1}$ are around $4.1 \times 10^{13}$ cm$^{-2}$, $1.9 \times 10^{13}$ cm$^{-2}$ and $1.3 \times 10^{13}$ cm$^{-2}$, respectively. This yields a total carrier density $n_S \approx 7.3 \times 10^{13}$ cm$^{-2}$, which is lower than that for the free surface ($1.26 \times 10^{14}$ cm$^{-2}$ in Ref [25] ; $2 \times 10^{14}$ cm$^{-2}$ in Ref. [26]).

**Figure 3** presents magnetotransport data for a series of Al//KTO samples. The temperature ($T$) dependence of the sheet resistance ($R_S$) reveals the presence of a 2DEG for all Al thicknesses with residual resistivity ratios in excess of 10. The low temperature $R_S$ shows a general decrease with increasing Al thickness (Fig. 3c). The Hall effect at 2 K (Fig. 3b) exhibits a nonlinear variation with the magnetic field, which suggests the presence of at least two types of carriers. Fitting the data with a two-band model yields the carrier densities $n_{S1}$ and $n_{S2}$ plotted in Fig. 3d and the corresponding mobilities $\mu_1$ and $\mu_2$ shown in Fig. 3e. For both bands, the carrier densities and mobilities tend to increase with Al thickness. The majority carriers (with density $n_{S1}$) have relatively low mobilities, in the 100-500 cm²/Vs range, while the minority carriers (with density $n_{S2}$) have mobilities reaching ~2000 cm²/Vs. The maximum total carrier density extracted from the Hall data reaches ~6×10¹³ cm⁻² and is thus slightly lower than that found in ARPES. This is possibly due to some reoxidation of the KTO when the samples are exposed to the atmosphere. However, one can ascribe the high-mobility minority carriers to the $E_{\beta}^{n=1}$ band and the lower-mobility majority carriers to the $E_{\alpha}^{n=1}$ and $E_{\alpha}^{n=2}$ bands. The apparent inconsistency between the high mobility and the larger effective mass of carriers from the $E_{\beta}^{n=1}$ band has previously been seen in STO 2DEGs and is ascribed to the larger distance of these carriers from the physical interface[27,28], which leads to longer scattering times $\tau$ (here in the range of $\tau_2 \approx 0.6$ ps, $vs$ $\tau_1 \approx 0.04$ ps for the low-mobility carriers from bands $E_{\alpha}^{n=1}$ and $E_{\alpha}^{n=2}$).

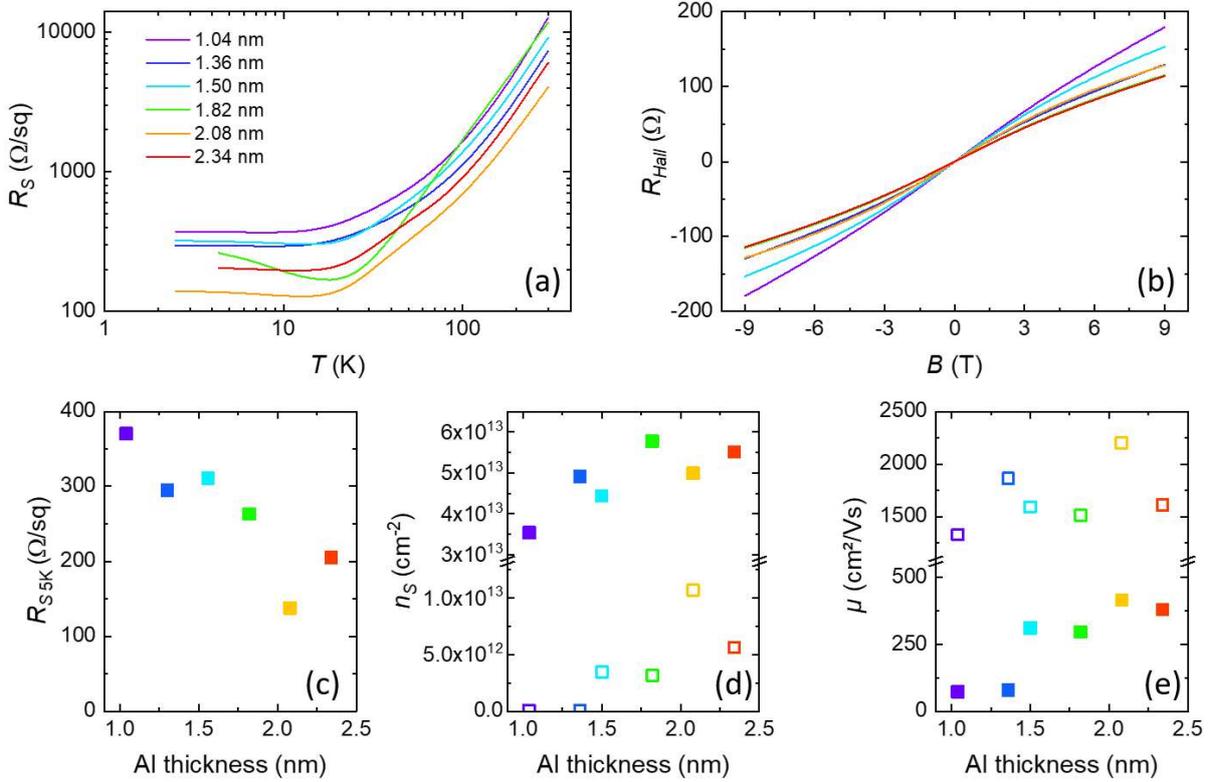

**Figure 3.** (a) Temperature dependence of the sheet resistance of Al//KTO samples with different Al thicknesses. (b) Hall effect at 2 K. (c) Sheet resistance at 5 K, (d) sheet carrier densities ($n_{s1}$: filled symbols; $n_{s2}$: open symbols), and (e) electron mobilities ($\mu_1$: filled symbols; $\mu_2$: open symbols)  as a function of the Al thickness.

To probe spin-charge conversion through SP-FMR experiments, we prepared AlO$_x$/NiFe/Al(0.9 nm)//KTO samples where the AlO$_x$-capped NiFe layer was grown in the same chamber and in the same vacuum cycle as the Al. The NiFe thickness was 20 nm but we also grew samples with a 2.5 nm thick NiFe layer. We performed magnetotransport measurements to extract the carrier densities and mobilities in the 2DEG[29,30], from which we found a total carrier density of about $1\times10^{14}$ cm$^{-2}$.

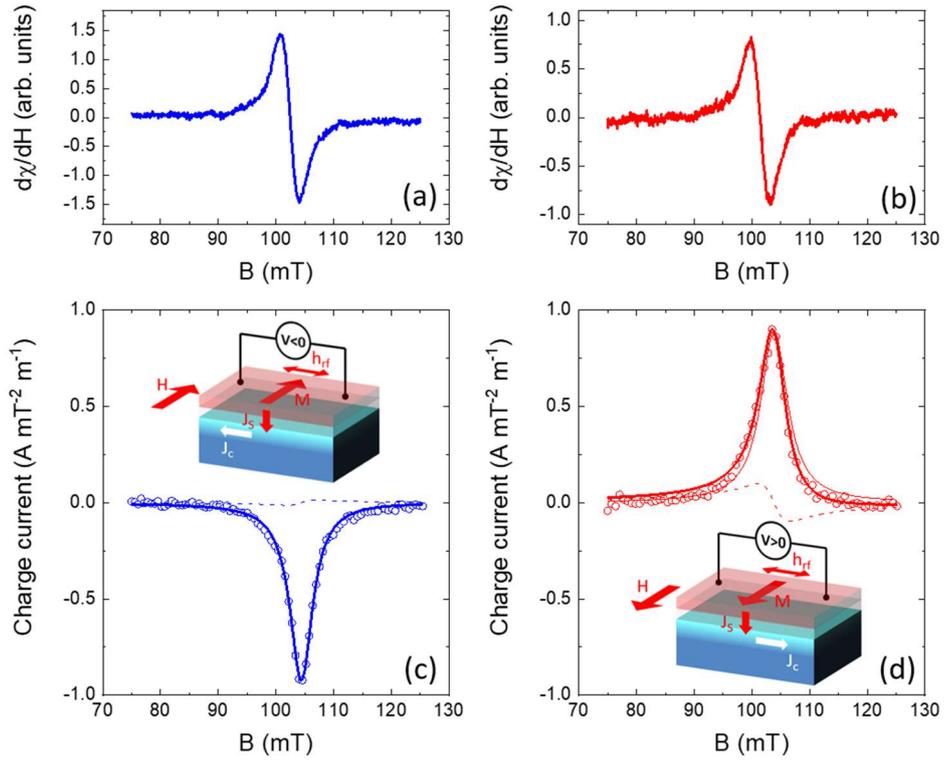

**Figure 4.** Derivative of the ferromagnetic resonance spectra at 10 K for negative (a) and positive (b) magnetic fields. Corresponding charge current produced transverse to the magnetization for negative (b) and positive (d) magnetic fields. The data in (b) and (d) are shown as symbols and the result of the fit as a thick solid line, while the symmetric and antisymmetric components are shown as thin solid and dashed lines, respectively.

**Figure. 4** shows the SP-FMR results. Fig. 4a and Fig 4b displays the FMR response of the NiFe layer for two opposite directions of the magnetic field. The resonance field is close to 100 mT, which is consistent with earlier results[8,11]. Fig. 4c and Fig 4d present the detected charge current, which exhibits a clear symmetric peak about the resonance field. As expected for spin-charge conversion, the signal changes sign when the magnetic field is reversed, i.e., when the magnetization and thus the spin polarization of the injected spin current are inverted.

For both configurations, the signal is largely dominated by a symmetric response due to the IEE, while the asymmetric component, which can arise from anisotropic magnetoresistance or planar Hall effect in the NiFe, is weak. We note that the sign of the conversion is opposite to that of LAO//STO at large negative gate voltages[8], which would correspond here to a negative Rashba coefficient. From the extra damping induced by the 2DEG with respect to the damping of a reference NiFe layer, the spin current density injected into the 2DEG can be calculated and used to estimate the spin-charge conversion efficiency $\lambda_{IEE} = \frac{J_C}{J_S}$ [4]. which we find here to be $\lambda_{IEE} \approx -3.5$ nm. This is comparable to what was measured in LAO/STO 2DEGs [8,31,32] (from 2 to $-6.4$ nm depending on the gate voltage) and is among the largest values reported to date, albeit lower than that for Al/STO ($\lambda_{IEE} \approx 30$ nm)[11]. We also note that the raw current produced is of the order of 40 nA, i.e., much higher than the value of 1 nA reported in thermal spin injection experiments[24].

To investigate charge-spin conversion – that has been little studied in oxide 2DEGs – we use magnetotransport and detect the unidirectional magnetoresistance[15,33,34]. This non-reciprocal phenomenon[35] is a consequence of the generation of a transverse spin density by the DEE. Its amplitude should vary linearly with both the current and the magnetic field; thus, it is often referred to as the bilinear magnetoresistance (BMR). Here, we have probed the BMR by measuring the 2DEG longitudinal resistance while rotating the magnetic field in the plane with respect to the current by an angle $\phi$ ($\phi$=0 is defined for the field parallel to the current). Together with the BMR, a quadratic MR (QMR), scaling with the square of the magnetic field, also arises. Remarkably, the ratio of their amplitudes ($A_{BMR}$ and $A_{QMR}$) allows extracting the Rashba coefficient of the system[15]:

$$\frac{A_{BMR}}{A_{QMR}} = \frac{2\pi\hbar}{eg\mu_B}\frac{\alpha_R}{E_F}\frac{J_C}{B}. \tag{1}$$

Here $\hbar$ is the reduced Planck constant, $e$ is the elementary charge, $\mu_B$ is the Bohr magneton and $g$ is the $g$-factor.

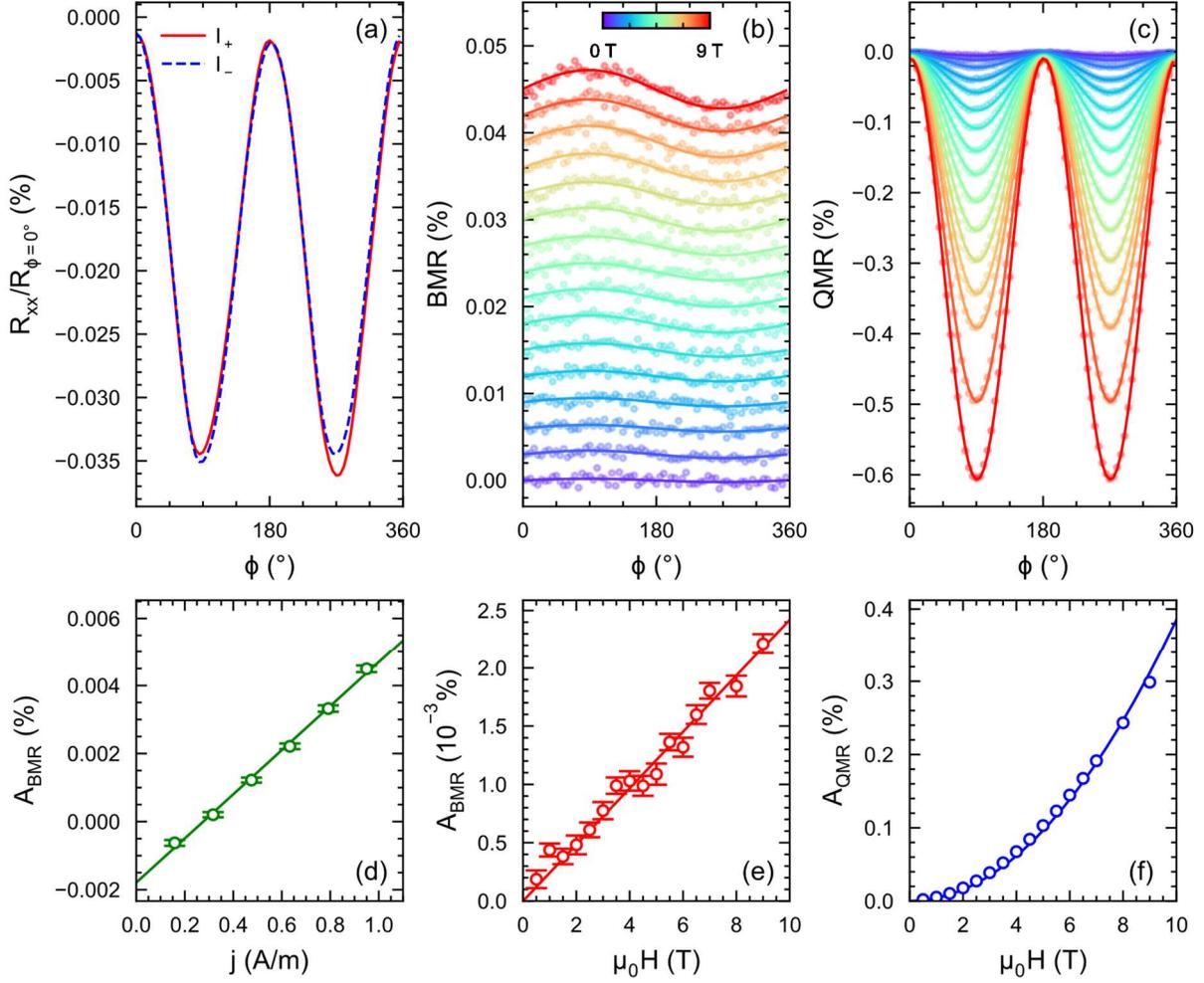

**Figure 5.** (a) Angle dependence of the normalized longitudinal resistance for two opposite currents ($J_C=\pm0.64$ A/m) and $B$=2 T. Angle dependence of the bilinear magnetoresistance (b) and the quadratic magnetoresistance (c) for increasing fields from 1 T to 9 T. (d) Dependence of the BMR amplitude with current at 9 T. Dependence of the BMR (e) and QMR (f) with magnetic field. All data at 2 K. In (b), (c), (e) and (f) $J_C$=0.64 A/m.

**Figure. 5a** shows the dependence of the magnetoresistance $\Delta R/R$ with the angle $\phi$ for positive and negative currents for a sample with 2.1 nm of Al at 2 T. The data are dominated by a $\cos 2\phi$ dependence (the QMR) but show slight shifts of opposite amplitudes at 90 and 270 degrees depending on the current sign, as a result of a $\sin \phi$ term (the BMR). We extract the BMR and QMR traces through the half-difference and half-sum of the curves measured at positive and negative currents, which are shown for increasing magnetic fields in Fig. 5b and Fig. 5c, respectively. The sign of the BMR is opposite to that found for LAO//STO[15] and, in the framework of Ref. [15], corresponds to a negative Rashba coefficient, which is consistent with the spin-pumping results. The amplitude of the BMR and QMR extracted from fits to the data are displayed in Fig. 5e and 5f, respectively. As expected, the BMR scales linearly with magnetic field, while the QMR scales quadratically. Fig. 5d shows the dependence of $A_{BMR}$ with the current, which is also linear, albeit with a small negative offset.

To exclude spurious thermal effects as the source of the BMR, we also measured the angular dependence of the longitudinal and transverse resistances using harmonic transport at a frequency $f$ (the QMR and the BMR appear in the $1f$ and $2f$ longitudinal signals, respectively) [34]. We found that the ratio of the transverse to the longitudinal signals was lower than the geometric factor of the Hall bar by at least a factor 3, which suggests that the contribution from thermal effects, if any, is very weak (see Supplementary Material).

Estimates of the Rashba coefficient $\alpha_R$ using Eq. (1) require the knowledge of the Fermi energy $E_F$ and the $g$-factor. As discussed earlier, the electronic structure of KTO 2DEGs comprises several bands with different Fermi energies and effective masses, which complicates the analysis. Nevertheless, due to their high mobility, electrons in the $E_\beta^{n=1}$ band carry more than 60% of the current. Because they are also expected to exhibit the largest Rashba coefficient[23], we can neglect contributions from the other bands to a good approximation. $\alpha_R$ also depends linearly of the value of the $g$-factor which, to the best of our knowledge, has not been measured in KTO. For the sake of comparison with STO data, we use the same value as that of Ref. [15] adapted from Ref. [10], namely $g$=0.5. We find $\alpha_R \approx -70$ meV.Å, i.e., two to three times higher (in absolute value) than the value found for STO 2DEGs. This value of the Rashba coefficient agrees well with the value extracted from weak-antilocalization data, $|\alpha_R| \approx 100$ meV.Å[23], and from Shubnikov-de Haas oscillations, $|\alpha_R| = 86$ meV.Å[36]. With $g$=2 we obtain $\alpha_R \approx -280$ meV.Å, compatible with the values computed for band $E_\beta^{n=1}$ [23]. Note that

in contrast to other techniques used to determine the Rashba coefficients, the BMR allows its sign to be determined.

We can now use these values of $\alpha_R$ to calculate $\lambda_{IEE} = {\alpha_R \tau}/{\hbar}$, which with $\tau$ =0.6 ps yields $\lambda_{IEE} \approx -6\text{-}25$ nm, depending on the value of $g$, which can be compared with the value of $-3.5$ nm extracted from SP-FMR. These values differ but fall within the same range. This discrepancy might be due to the over-simplification of the BMR analysis or to the fact that different carrier densities, and thus electronic structure, appear between the samples used for SP-FMR and BMR measurements.

In summary, we have synthesized a new Rashba two-dimensional electron gas by depositing Al at room temperature on commercial (001)-oriented KTaO₃ single-crystal substrates. As for the Al//STO system, the deposition of Al reduces the KTO, which promotes the formation of the interfacial 2DEG. The carrier densities, mobilities, and sheet conductivities are found to increase systematically with the Al thickness, while ARPES measurements provide evidence of a complex multiband structure, reminiscent of that of KTO surfaces. By using spin-pumping FMR and unidirectional magnetoresistance experiments, we demonstrated very efficient spin-charge and charge-spin conversion. We obtained consistent results between spin-charge and charge-spin conversion experiments and extracted a negative Rashba coefficient in the range of 70 to 280 meV.Å, which is significantly higher than that found in STO 2DEGs. The spin-conversion efficiency $\lambda_{IEE}$ is among the highest reported in the literature[4–6,8,11] to date, and is an order of magnitude larger than that with transition metals such as Pt (comparing $\lambda_{IEE}$ with the product of the spin Hall angle and spin diffusion length, i.e. $\theta_{SHE} \times l_{sf}$). We suggest that $\lambda_{IEE}$ could probably be enhanced substantially by increasing the momentum relaxation time, which appears feasible in light of the very high mobilities (>10000 cm²/Vs) reported in bulk or irradiated KTO[17,21]. Accordingly, KTO 2DEG might represent an interesting candidate for the readout unit of MESO transistors[37]. Further experiments should aim at determining the $g$-factor of this material, along with investigating spin-charge interconversion as a function of gate voltage and at higher temperatures. Combined with the very recent reports of a superconducting state at (111)- and (110)-oriented KTO interfaces[38–40], our findings of a large Rashba coefficient in this system also offer interesting perspectives for topological superconductivity and Majorana physics.

## Experimental Section

***Preparation of KTO substrates:*** Single-crystal KTO (001) substrates (5 mm × 5 mm × 0.5 mm, one-side polished with miscut angles < 0.1°) were purchased from SurfaceNet GmbH. The as-received substrates were cleaned by sonicating in deionized water, acetone and isopropanol and subsequently dried with nitrogen. This process was repeated until the cleanliness of the surface was confirmed by AFM. The cleaned substrates were inserted into a UHV system for metal deposition by magnetron sputtering and in-situ XPS measurements.

***Al deposition by sputtering:*** The Al deposition was performed at room temperature in a commercial dc magnetron sputtering system PLASSYS MP450S with a base pressure of 9 x $10^{-8}$ mbar. The deposition rate was derived by means of x-ray reflectometry (XRR) on thicker samples grown under the same conditions. Ar gas flow and the current intensity were fixed to 5.2 sccm and 30 mA, respectively. The pressure of the chamber during deposition was 5.3 ± 0.2 x $10^{-4}$ mbar and the plasma power was 10 W. The NiFe was also deposited by dc magnetron sputtering and capped with 1.5 nm of Al, which transformed into $AlO_x$ after exposure to air.

***Al deposition by molecular beam epitaxy:*** After pre-annealing the KTO at 200°C for 1 h in UHV, we grew 2 Å of Al at room temperature at $7.10^{-10}$ mbar using a Knudsen cell heated to 1000°C at a growth rate of 0.011 Å/s.

***XPS measurements:*** The XPS measurements were performed at room temperature using an Omicron NanoTechnology GmbH system with a base pressure of 5 x $10^{-10}$ mbar, using a Mg $K_\alpha$ source ($h\nu = 1253.6$ eV) operating at 20 mA and 15 kV. The spectra were obtained at a pass energy of 20 eV. In-situ XPS measurements were performed before and immediately after the deposition of Al. The spectral fits were carried out using CasaXPS (CasaSoftware Ltd.).

***LEED measurements***: LEED diffraction patterns have been measured using an Omicron SPECTALEED. All images have been acquired at room temperature.

***ARPES measurements***: ARPES experiments were performed at the CASSIOPEE beamline of the SOLEIL synchrotron light source (Saint-Aubin, France). The CASSIOPEE beamline is equipped with a Scienta R4000 hemispherical electron analyser with angular acceptance of $\pm15°$ (Scienta Wide Angle Lens). All the experiments were performed at room temperature. The angle and energy instrument resolutions were 0.25° and 12 meV, respectively. The incident photon beam was focused into a 50-$\mu$m spot (in diameter) on the sample surface. All ARPES measurements were performed with a linearly-polarized photon beam and at the photon energy $h\nu$= 30 eV. The collected data were normalized by the intensity background of the electron analyzer and smoothed using an averaging filter.

***Magnetotransport properties measurements:*** The samples were measured with a Dynacool system from Quantum Design after bonding with Al wires in the Van der Pauw configuration. During the transport measurements of AlO$_x$/NiFe/Al//KTO samples, the NiFe and 2DEG signal were probed in parallel[41]. These contributions were separated in the following way. For the longitudinal configuration, $R_{xx}$, two resistances in parallel were measured so that the resistance of the 2DEG alone is given by

$$R_{2DEG} = (R_M \times R_{Total}) / (R_M - R_{Total})$$

For the transverse configuration, $R_{xy}$ (Hall resistance), besides $R_M$ and $R_{2DEG}$ the Hall voltages generated in each layer must be also considered. These circuits can be simplified using Millman's theorem[42] so that the Hall resistance of the 2DEG alone $R_{H,2DEG}$ is given by

$$R_{H,2DEG} = R_{H,Total} \times ((R_M / R_{Total}) + 1)^2 - R_{H,M} \times (R_{2DEG} / R_M)^2$$

The 2DEG contribution was then fitted with a standard two-band model in order to extract carrier densities and mobilities.

***Unidirectional magnetoresistance measurements***: Angle-dependent magneto-transport experiments were performed on a 1.58 x 10 mm slab of Al(2.1 nm)//KTO. Electrical contacts were made by Al wedge bonding. For dc measurements, a current was applied between the extremities of the slab and the longitudinal resistance was measured between two lateral contacts separated by 5530 µm, using a Keithley 2400 sourcemeter. Harmonic measurements were performed in a Hall bar device (10 x 100 µm) by injecting an a.c. current (Keithley 6221) at $f$ = 3 kHz and demodulating the longitudinal and transverse voltages with a lock-in amplifier (Zurich Instruments HF2LI). The unidirectional terms appear in the second harmonic ($2f$ signal). In-plane angle dependent magnetoresistance measurements were performed by rotating the sample inside a Dynacool PPMS, at constant applied magnetic field, temperature and current. Angular scans (Fig. 5a) are plotted after subtraction of a linear drift and a sinusoidal background ascribed to out-plane residual magnetic field.

***Spin-pumping ferromagnetic resonance***: The spin-to-charge interconversion in the 2DEG at the surface al KTO was performed using the spin-pumping ferromagnetic resonance technique[43] in a cavity at 10 K on an AlOx/NiFe(20 nm)/AlOx//KTO sample. A d.c. and radiofrequency field at 9.7 GHz were applied to the system. The r.f. frequency was kept constant while the amplitude of the d.c. field was swept around the resonance field. At FMR, the magnetization of the NiFe film precesses uniformly around the direction of the d.c. field, creating a spin accumulation that leads to the injection of a pure spin current into the 2DEG at the AlOx/KTO interface. The injected spin current is subsequently converted into a charge current in the 2DEG by the IEE. Fig. 4 shows the charge current produced by unit of applied r.f. power at a gate voltage of -120 V. This current possesses a symmetric and an antisymmetric component. The antisymmetric part corresponds to rectifications effect such as the planar Hall Effect[44]. The symmetric component is due to the spin to charge interconversion by the IEE in the 2DEG. The value of the injected spin current is computed by measuring the magnetization ($M_s$ = 817 kA m$^{-1}$), the g-factor ($g$ = 2.1) enhancement of the damping between a reference 20 nm-thick NiFe film ($\alpha_0 = 6.36 \cdot 10^{-3}$) and the AlOx/NiFe(20 nm)/AlOx//KTO sample ($\alpha_0 = 6.94 \cdot 10^{-3}$) [11]. The ratio of the spin and charge currents gives the Edelstein length.

**Acknowledgements**

L.M.V.-A., J.B. and S.V. contributed equally to this work. The authors thank Marc Gabay, Annika Johansson, Börge Göbel and Ingrid Mertig for useful discussions and Joo-Von Kim for a careful reading of the manuscript. This work received support from the ERC Advanced grant n° 833973 "FRESCO", the QUANTERA project "QUANTOX", the Laboratoire d'Excellence LANEF (ANR-10-LABX-51-01), the ANR projects TOPRISE (ANR-16-CE24-0017) and QUANTOP (ANR-19-CE47-0006-01) and Intel's Science and Technology Center - FEINMAN. F. Trier acknowledges support by research grant 37338 (SANSIT) from VILLUM FONDEN.

**WILEY-VCH**

**Table of contents**

Two-dimensional electron gases formed by sputtering a thin film of Al onto $KTaO_3$ crystals are used to interconvert charge and spin currents with a high efficiency, thanks to the large Rashba spin-orbit coupling in the electron gas. These results are relevant for the integration of quantum oxide materials into spintronics devices.

*Luis M. Vicente-Arche, Julien Bréhin, Sara Varotto, Maxen Cosset-Cheneau, Srijani Mallik, Raphaël Salazar, Paul Noël, Diogo Castro Vaz, Felix Trier, Suvam Bhattacharya, Anke Sander, Patrick Le Fèvre, François Bertran, Guilhem Saiz, Gerbold Ménard, Nicolas Bergeal, Agnès Barthélémy, Hai Li, Chia-Ching Lin, Dmitri E. Nikonov, Ian A. Young, Julien Rault, Laurent Vila, Jean-Philippe Attané and Manuel Bibes*

**Spin-charge interconversion in $KTaO_3$ two-dimensional electron gases**

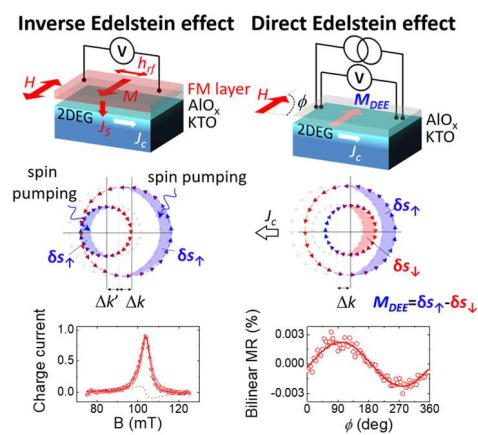

# Supplementary Information

## Spin-charge interconversion in KTaO$_3$ two-dimensional electron gases

*Luis M. Vicente-Arche, Julien Bréhin, Sara Varotto, Maxen Cosset-Cheneau, Srijani Mallik, Raphaël Salazar, Paul Noël, Diogo Castro Vaz, Felix Trier, Suvam Bhattacharya, Anke Sander, Patrick Le Fèvre, François Bertran, Guilhem Saiz, Gerbold Ménard, Nicolas Bergeal, Agnès Barthélémy, Hai Li, Chia-Ching Lin, Dmitri E. Nikonov, Ian A. Young, Julien Rault, Laurent Vila, Jean-Philippe Attané and Manuel Bibes\**

Luis. M. Vicente-Arche, Julien Bréhin, Dr. Sara Varotto, Dr. Srijani Mallik, Dr. Diogo Vaz[♣], Dr. Felix Trier[♥], Suvam Bhattacharya[♦], Dr. Anke Sander, Prof. Agnès Barthélémy & Dr. Manuel Bibes
Unité Mixte de Physique, CNRS, Thales, Université Paris-Saclay, 1 avenue Augustin Fresnel, 91767, Palaiseau France
E-mail : manuel.bibes@cnrs-thales.fr
[♣] Present address: CIC nanoGUNE BRTA, Tolosa Hiribidea, 76, 20018 Donostia - San Sebastian, Spain
[♥] Present address: Department of Energy Conversion and Storage, Technical University of Denmark, Fysikvej, Building 310, 2800 Kgs. Lyngby, Denmark
[♦] Present address: Laboratoire de Physique des Solides, CNRS, Université Paris-Saclay, 91405 Orsay, France

Maxen Cosset-Cheneau, Dr. Paul Noël[§], Dr. Laurent Vila, Dr. Jean-Philippe Attané
Université Grenoble Alpes, CEA, CNRS, Grenoble INP, SPINTEC, 38000 Grenoble, France
[§] Present address : Dept. of Materials, ETH Zürich, Hönggerbergring 64, 8093 Zürich, Switzerland

Raphaël Salazar, Dr. Patrick Le Fèvre, Dr. François Bertran, Dr. Julien Rault
Synchrotron SOLEIL, L'Orme des Merisiers, Saint-Aubin, BP 48, 91192 Gif-sur-Yvette Cedex, France

Guilhem Saiz, Gerbold Ménard, Nicolas Bergeal
Laboratoire de Physique et d'Etude des Matériaux, ESPCI Paris, Université PSL, CNRS, Sorbonne Université, Paris, France.

Hai Li, Chia-Ching Lin, Dmitri Nikonov, Ian Young
Components Research, Intel Corp., Hillsboro, OR 97124, USA

**X-ray photoelectron spectroscopy of Al 2p levels**

We have measured the Al 2p state after Al deposition for a representative sample with 21 Å of Al, in situ and ex situ (after exposure of the sample to the atmosphere) see Fig. S1. The results are quantitatively quite similar to those reported in Ref. 12: after exposure to the air, in this thickness range, the Al is almost completely oxidized. The remaining metallic Al regions most probably do not percolate as they do not appear to contribute to the transport (this can be checked by measuring the Hall signal at room temperature: if a continuous metallic region is present, a large part of the current will flow through it and it will thus contribute to the Hall effect with a very small slope, corresponding to its large carrier density ; in all the samples discussed in the paper, the slope of the Hall effect at room temperature is on the order of 10 $\Omega$/T, similar to the slope at low temperature).

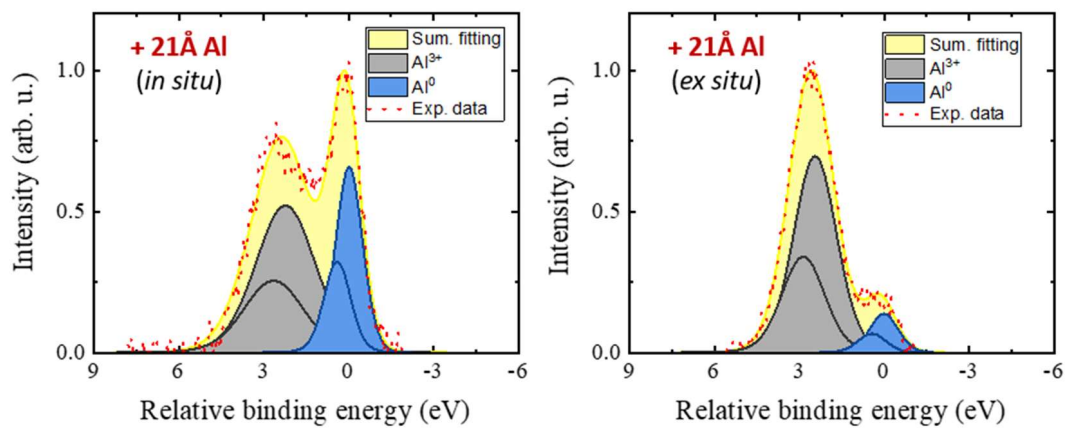

**Fig. S1.** *Al 2p XPS spectra of a Al (21 Å)/KTO sample before (left) and after (right) exposure to the atmosphere. While a significant fraction of metallic Al remains in the in situ data (36%) it is reduced to only 12% ex situ.*

**Angle dependent transport measurements**

In the raw R vs $\phi$ data we identify the presence of a small linear drift and of a sinusoidal background ascribed to a slight spurious Hall signal produced by a tiny out-of-plane magnetic component (caused by a slight sample misalignment), cf Fig. S2a. We have corrected the data for these two spurious contributions, cf Fig. S2b. Note that the correction does not affect the amplitude of the BMR or QMR terms.

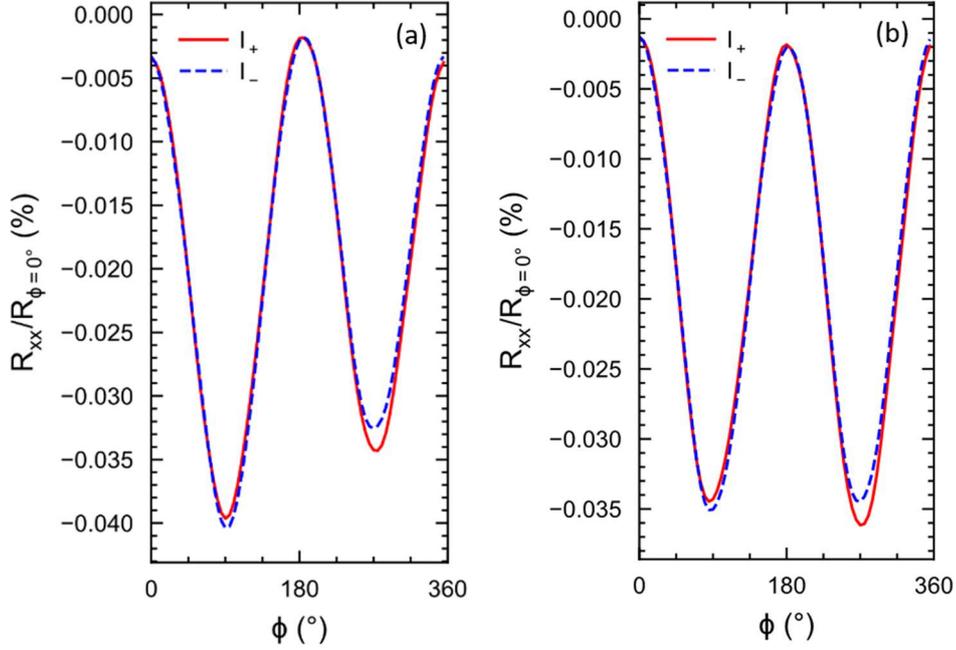

**Bilinear magnetoresistance and thermoelectric effects**

In this work we ascribed the bilinear magnetoresistance to the inversion symmetry breaking and Rashba SOI. In the literature, the Rashba-driven BMR can be mistaken for nonreciprocal thermal effects due to a vertical temperature gradient, appearing due to unintentional heating by the applied current (e.g. Nernst effect [1], nonlinear Seebeck effects [2] or spin Seebeck and anomalous Nernst in case of ferromagnetism). Here, since we work with a 2DEG, we do not expect a thermal gradient to develop along the quantization direction z. However, to check for the possible contribution from thermal effects, we did measure the angle dependence of the longitudinal and transverse resistances using harmonic transport at a frequency $f$ (the QMR and the BMR appearing in the 1$f$ and 2$f$ longitudinal signals, respectively) in Hall bar devices. If thermal effects are present, a voltage will develop at 2$f$ along the longitudinal and transverse direction, in proportion of the length and width of the Hall bar, whose ratio was 10 in our device.

The angle sweeps in Figure S3 show that the ratio of the transverse to the longitudinal 2*f* signals terms is at least a factor 30, i.e. much higher than the geometric ratio. This indicates a weak contribution to the BMR signal from thermoelectric effects, if any.

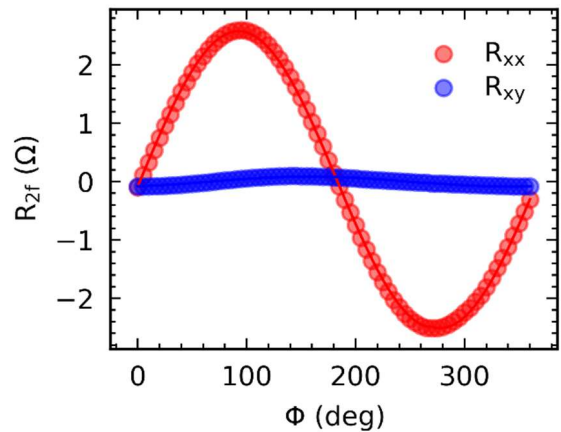

**Fig. S2.** *Angle dependence of the second harmonic longitudinal ($R_{xx}$) and transverse ($R_{xy}$) resistance at B = 4 T, measured in a Hall bar KTO/Al device with width W = 10 μm and length L = 100 μm (current density J = 50 A/m, frequency f = 3 kHz).*